\documentclass[aps,prb,twocolumn]{revtex4}

\usepackage{graphics,amsmath}

\usepackage{graphicx}

\begin{document}

\title{Optical and 
spectral properties of quantum domain-walls in the\\ generalized
  Wigner lattice}  
 
\author{S. Fratini$^{1}$}
\author{G. Rastelli$^{1,2}$}
 
\affiliation{$^{1}$Institut N\'eel,
D\'epartement Mati\`ere Condens\'ee et Basses Temp\'eratures 
- CNRS\\ 
BP 166, F-38042 Grenoble Cedex 9, France}
 
\affiliation{$^{2}$Istituto dei Sistemi Complessi, CNR-INFM\\ 
v. dei Taurini 19, 00185 Roma, Italy}

\begin{abstract}
We study the spectral properties of a  system of electrons interacting
through long-range Coulomb potential  on a one-dimensional chain.  
When the interactions dominate over the electronic bandwidth,
the charges arrange in an ordered configuration that minimizes the
electrostatic energy, forming Hubbard's generalized Wigner lattice. 
In such strong coupling limit, the low energy excitations are quantum
domain-walls that behave as  fractionalized charges, and can be bound
in excitonic pairs.  
Neglecting higher order excitations,
the system properties are 
well described by an effective Hamiltonian in the 
subspace with one pair of domain-walls, which can be solved exactly.
The optical conducitivity $\sigma(\omega)$ and 
the spectral function $A(k,\omega)$ can be calculated analytically, 
and reveal unique features of the unscreened Coulomb interactions that
can be directly observed in experiments.
\end{abstract}

\date{\today}

\maketitle

\section{Introduction}

Quantum many-body systems often exhibit complex behavior, 
arising from the strong interactions between the individual degrees
of freedom.  
In some cases, the exotic physical properties can be simply  
explained in terms of suitable renormalized collective 
excitations. Well-known examples are  the charge-spin separation 
in the Luttinger liquid with gapless bosonic excitations, \cite{Voit} solitons  
on a CDW condensate, \cite{Rice} and fractional charges in
quantum Hall systems.\cite{Laughlin}

Another system that exhibits emerging complex behavior is the generalized 
Wigner lattice (GWL), defined as the classical  charge pattern that minimizes the 
Coulomb repulsion on a discrete lattice. 
The GWL  was introduced by Hubbard\cite{Hubbard78} to explain
the charge ordering observed in the TTF-TCNQ organic salts, and 
has since  been invoked in several classes of narrow-band 
quasi-one-dimensional compounds.\cite{Itou,Abbamonte,Horsch}
Increasing the electron bandwidth
smears the classical charge distribution, 
and eventually drives the system towards a small amplitude 
charge density wave.\cite{Valenzuela} 

However, in the strongly interacting limit, where the bandwidth is 
much smaller than the electrostatic repulsive energy, the quantum charge 
distribution remains very close to the classical configuration. In this limit,
the low-lying excitations are pairs of 
domain-walls (kinks and anti-kinks), that carry fractional charge.\cite{Hubbard78,SynthMet,Mayr}
It follows that the low energy properties of the system can be
determined by solving the problem of two interacting domain-walls, 
which is equivalent to the Coulomb problem on a one-dimensional chain.
Since kinks and anti-kinks have opposite charges,
the lowest lying excitations of the GWL  will be bound pairs, 
followed by a continuum of unbound domain-walls.
Restricting to the subspace with only one kink/anti-kink pair, 
the quantum melting of the GWL was estimated  in Ref.\cite{SynthMet} 
as the point where the gap in the excitation spectrum vanishes.
The result is in agreement with previous variational estimates \cite{Valenzuela} 
as well as with more recent exact diagonalization data, \cite{Mayr} indicating that the
single pair approximation captures the essential physics of the GWL phase. 
On the other hand, it
clearly breaks down close to the quantum melting
transition,  where a proliferation of domain-walls is expected.

In this paper, we take advantage of the exact solution of the discrete
Coulomb problem given in Refs.\cite{Gallinar84,Kvitsinski} to 
evaluate the optical conductivity  of the 
generalized Wigner lattice at simple commensurate fillings $n=1/s$.
The formation of excitons manifests 
through the emergence of a series of sharp peaks, 
followed by a strongly asymmetric absorption band due to 
continuum of scattering states.  
Remarkably, both the discrete peaks and the asymmetry of the 
absorption continuum are direct consequences of 
the long-range interactions among the electrons in the
original model, that disappear when the Coulomb interactions
are replaced by short-range potentials.

An analogous formalism is used to calculate the spectral function $A(k,\omega)$.
However, since the addition or removal of an electron to a GWL at filling 
$n=1/s$ is equivalent to the creation of $s$ domain-walls,\cite{SynthMet} 
the single pair approximation only applies to the special case $n=1/2$.
The  interactions being repulsive, because the domain-walls have equal
charge, the low-lying excitations in the spectral function  
are unbound scattering states, while anti-bound states appear  
as a set of dispersive quasi-particle peaks above the continuum.\cite{Daghofer}

The paper is organized as follows. In Sec. II we introduce the
one-dimensional model for spinless fermions with long-range interactions, 
and solve it in the narrow-band regime, restricting to the subspace with one
pair of domain-walls.
In Sections III and IV we use this analytical solution to 
calculate respectively  the optical conductivity   
and  the single particle spectral function. The results are briefly
discussed in Sec. V in connection with existing experimental work.

\section{Model and Single pair approximation}

We consider the following Hamiltonian for fermions
on a linear chain, interacting through long-range Coulomb forces: 
\begin{equation}
\label{hamiltonian}
H = -t \sum_{i} 
\left( {c}^{\dagger}_i {c}_{i+1} + 
{c}^{\dagger}_i {c}_{i-1}  \right) + 
\frac{V}{2} 
\sum_{i \neq j} 
\frac{
\left({n}_i - n \right) 
\left( {n}_j - n \right) 
}{\left| i -  j \right|} 
\end{equation}
Here $t$ is the nearest-neighbor hopping amplitude and
$V$ sets the  energy scale of 
the long-range repulsion, $V_{m}=V/|m|$.  
$\{ {c}^{\dagger}_i,{c}_i \}$
are, respectively, the fermionic creation and annihilation operators,  
${n}_i={c}^{\dagger}_i {c}_i$ 
is the occupation number at site $i$ and $n$ is the average charge per
site.  The spin degrees of freedom are explicitly neglected, corresponding to
the limit of large on-site repulsion $U$. The lattice parameter is
set to unity.

Since we are interested in  the strong coupling regime, it is useful
to start by describing the classical solution, obtained for $t=0$. 
At simple commensurate fillings  $n=1/s$, 
the ground state takes one of the $s$ 
equivalent configurations that minimize the electrostatic
energy, with one electron on every $s$-th site. 
\footnote{The electrostatic energy  {\sl per particle}  
in the classical state is:
\[
\mathcal{E}_{GS} = \frac{s-1}{s^2} \gamma + \frac{2}{s^3} 
\sum^{s-1}_{m=1} m \left[ \Psi\left( \frac{m}{s} \right) + 
\frac{\pi}{s} \cot \left(\frac{\pi m}{s} \right)
\right]
\]
where $\gamma$ is the Euler constant and $\Psi(x)$ is the Digamma function.
} 
The excited configurations of lowest energy  
correspond  to the inclusion of a domain with a different (shifted)
ground state, delimited by a  kink and an anti-kink.\cite{SynthMet}
The energy cost to form  a classical domain of finite lenght $sd$
($d$ is the  number of perturbed unit cells)  can be calculated as
\begin{equation}
   \Delta_s(d)=\sum_{p=1}^d \sum_{m=p}^\infty
   \left(V_{sm+1}+V_{sm-1}-2V_{sm}\right).
\label{eq:kk-energy-general}
\end{equation}
It defines an effective interaction between 
the domain-walls, and can  be evaluated exactly in terms of Digamma
functions. However, a very good approximation is obtained by 
expanding in powers of $1/sd$ and keeping only the leading term 
\begin{equation}
 \label{eq:pot} 
\Delta_s \left( d \right) \simeq 
\Sigma_s  - \frac{V}{s^2} \frac{1}{s d} .
\end{equation}
The above expression can be directly interpreted as  
the  Coulomb attraction between two defects of charge 
$\pm 1/s$ at distance $sd$, with
$\Sigma_s= \frac{V}{s}\left[1-\frac{\pi}{s}\cot
\left(\frac{\pi}{s}\right)\right]$ the energy cost for 
creating two isolated defects at infinite distance. 
The approximate result Eq.(\ref{eq:pot}) is quite accurate even at short
distances. 
\footnote{The effective potential can be evaluated to:
$$
\Delta_s \left( d \right)   =  \Sigma_s + \frac{V}{s^2} \left[ 
\Psi \left( d - 1/s \right)  - \Psi \left( d + 1/s \right)
\right]
$$
$$ 
+ V \frac{d}{s} \left[  2 \Psi \left( d \right)  -
\Psi \left( d + 1/s \right)  - \Psi \left( d - 1/s \right) 
\right] 
$$
The largest discrepancy 
 occurs for  $s=2$ and  $d=1$, where the above expression gives
 $\Delta_2(1)  \simeq 0.386 V$ while the value for the pure Coulomb law is 
$0.375 V$.}
Unless otherwise specified, we shall neglect the higher order terms (dipolar,
multipolar interactions)  in the following. 

According to the above discussion, the gap in the excitation spectrum 
is given classically by  ${\Delta}_s(1)$. 
Actually, this can be viewed as the effective coupling parameter of
the model, that determines the stability of
the classical charge pattern against quantum fluctuations, so that the
melting of the GWL is expected when  $t \gtrsim {\Delta}_s(1)$.\cite{Valenzuela}
Since the gap scales as $\sim V/s^3\sim V n^3$, we see that 
for a given value of the hopping
integral $t$, the formation of a GWL becomes unfavorable 
at very low fillings. 
In this case an ordinary Wigner crystal phase is more likely
realized, where the electron wavefunctions are spread on lengthscales much
larger than the lattice spacing.\cite{Valenzuela} 
A similar argument shows that the GWL phase is unfavored when moving
away  from the simple commensurate 
fillings $n=1/s$. Indeed, at generic rational  fillings $n=r/s$,  
it is easily shown that the gap  still scales as $\sim V/s^3$ (i.e. 
it is governed by the periodicity $s$ of the classical pattern), which
can be arbitrarily small regardless of the value of $n$.\cite{Slavin}
In the following, we shall therefore  
restrict to the special fillings $n=1/s$, where the physics of the 
GWL is more relevant.

\bigskip

Let us denote  $|m, d \rangle $ the classical state 
corresponding to a kink/antikink pair of length
$sd$, with center of mass at $m$. 
The total two-body wavefunction in the quantum case can be written as
\begin{equation}
  \label{eq:wavef}
  \Psi_{K,\lambda}=\sqrt{\frac{s}{2L}}
  \sum_{m=-L/s}^{L/s}\sum_{d=-L/2s}^{L/2s}
  e^{iKm} \varphi_{K,\lambda}(d) |m,d \rangle
\end{equation}
where the center of mass of the pair moves freely with momentum $K$, while the 
radial part  obeys the following discrete Schr\"odinger equation:\cite{SynthMet,Barford}
\begin{equation}
  \label{eq:disc-Schr} 
  -2t(K) [ \varphi(d+1)+ \varphi(d-1)]+
\left[   \Sigma_s  - \frac{V/s^3}{d} - E \right]   
\varphi (d) =0
\end{equation}
with the  boundary condition   $\varphi(0) =0$.  
The factor $t(K) =t \cos (Ks/2)$ reflects the fact that at each
electron hopping event, the center of mass of the defect pair 
moves by $\pm s/2$ lattice sites.   
The discrete hydrogen problem  on the semi-infinite chain 
admits an exact analytical  solution\cite{Gallinar84,Kvitsinski} in 
terms of Gauss-hypergeometric functions (or, equivalently, of
incomplete beta functions). 
The excitation spectrum consists, as in the ordinary hydrogen atom,  
of an {\sl infinite} series of bound states (kink/anti-kink excitons),
followed by a continuum of scattering states.  
Observing that the two subspaces of
different polarity $d>0$ and $d<0$  are disconnected
when only nearest-neighbor electron hopping is allowed,\cite{Mayr} 
the solution of the Coulomb problem Eq.(\ref{eq:disc-Schr}) 
on the infinite chain is obtained straightforwardly  by constructing 
combinations of even or odd parity of the wavefunctions on the half-chain.

\medskip 
\paragraph{Domain-wall excitons.}
The bound states have energies:
\begin{equation}
  \label{eq:bound} 
  E_{K,n}= \Sigma_s -  \sqrt{ 16 t(K)^2+\left( \frac{V}{ s^3
        (n+1)  }\right )^2}
\end{equation}
with $n=0,1,2,\ldots,\infty$. Introducing the parameter 
$\sinh(k_n) = {V }/{[ 4 s^3 t(K) (n+1)]}$,
the normalized radial wavefunctions
read:
\begin{equation}
\varphi_{K,n}(d) = A_n \; d \; e^{-k_n d} \;
{}_2F_1(1-d,-n;2;1-e^{2k_n}) \label{wavedisc}
\end{equation}
where ${}_2F_1$ is one of the Gauss hypergeometric functions and $A_n$
is a normalization constant:
\begin{equation}
A_n  = 2 \sinh(k_n)  \sqrt{\tanh(k_n)} \;
e^{-n k_n}  \label{normdisc}
\end{equation}

The  exciton radius, defined as 
$r_n=\sum^{\infty}_{d=1} d \; {\left| \varphi_{K,n}(d)  \right|}^2$, can also 
be  evaluated: \cite{Kvitsinski}
\begin{equation}
r_n = \left( n + 1 \right) 
\frac{2+\cosh (2k_n) }{\sinh(2k_n) }
\end{equation}
In the strong coupling regime, $r_n \approx (n+1)$ and the first 
bound state $(n=0)$
is strongly localized on one site. At large quantum numbers the radius
increases as $r_n \simeq (6 s^3t/V) (n+1)^2 $ approaching the free
scattering states.

\medskip
\paragraph{Scattering states.}
The energy of the scattering states   
is given by the sum of the energies of the individual domain-walls, with  
$k$ the relative momentum of the two-body system:
\begin{equation}
  \label{eq:continuum}
  E_{K,k}= \Sigma_s  - 2 t \left [\cos(Ks/2+k)+ \cos (Ks/2-k) \right].
\end{equation}
The corresponding wavefunction  is:\cite{Kvitsinski}
\begin{equation}
  \label{eq:scatt}
  \varphi_{K,k} (d)=  B^{-1}(k) d e^{ikd}
  {}_2F_1(1-d,1-i\eta(k);2;1-e^{-2ik}) 
\end{equation}
where $\eta(k)= V/[2s^3 t(K) \sin k]$.   The normalization factor
 is determined by the asymptotic
behavior at large $d$, and is given by
\begin{equation}
  \label{eq:norm}
  B^{-2}(k)=\frac{4s}{L}\frac{\pi\eta(k) }{\sinh [\pi\eta(k)] } \sin^2 (k) \ e^{-(2k - \pi)\eta(k)}
\end{equation}
in the case of an open chain of length $L/2s$.

\medskip

\begin{figure}[htbp]
  \resizebox{8cm}{!}{\includegraphics{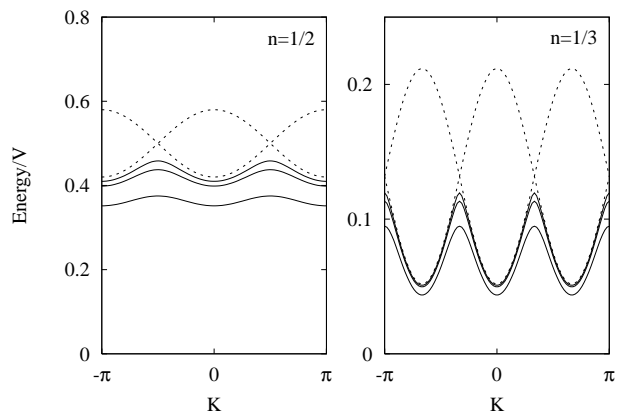}}
  \caption{Excitation spectrum vs. center of mass momentum $K$ 
in the single pair subspace for $t/V=0.02$  at fillings $n=1/2$ (left) and $n=1/3$ (right). 
The full lines correspond to the 3 lowest bound states, while the
dashed lines indicate the boundaries of the kink/anti-kink continuum.}
  \label{fig:spectrum}
\end{figure}

The excitation spectrum is illustrated in Fig. \ref{fig:spectrum} for $t/V=0.05$ at
fillings $n=1/2$ and $n=1/3$.
The gap in the excitation spectrum 
is obtained from Eq. (\ref{eq:bound}) setting  
$K=0$ and $n=0$:
\begin{equation}
\label{eq:GAP_opt} 
\Delta^{opt}_s= \Sigma_s - 4 t \sqrt{ 1+\left( \frac{V}{4t s^3  }\right )^2}.
\end{equation}
This value is lower than the classical value
$\Delta_s(1)$ because of the energy gain due to charge delocalization. 
The condition $\Delta^{opt}_s  
\rightarrow 0$ can be used to locate the melting of the GWL as the
region where charge defects are expected to proliferate.
\cite{SynthMet} For the quarter-filled case $s=2$, the estimate
$t/V\simeq 0.12$ is in good agreement with the numerical results of 
Ref.\cite{Mayr}.

\section{Optical conductivity}

The finite frequency absorption at $T=0$ can be expressed in terms of the
current-current correlation function through the Kubo formula: 
\begin{equation}
  \label{eq:Kubo}
  \sigma(\omega)=\frac{1}{L \omega} 
  \mbox{Im}   \sum_{K,\lambda} \frac{|\langle \Psi_{K,\lambda} |
  \hat{j} |  \psi_{GS}\rangle |^2 }{\omega -E_{K,\lambda} +i 0^+}
\end{equation}
where the sum runs over all eigenstates,  and the ground state energy
is set to $0$.   For nearest-neighbor hopping, 
the current operator on the lattice is defined as 
$\hat{j}= i t \sum_i (c^+_i c_{i+1}-c^+_{i+1}c_i)$. The above
dimensionless expression should be multiplied by a prefactor
$\sigma_0= e^2a^2/\hbar v$ to restore the proper units, 
where $a$ is the intersite distance on the
chain and $v$ is the volume of the unit cell. 

Approximating $|\psi_{GS}\rangle$ in the GWL phase at fillings $n=1/s$ with the
classical state $|  \psi_\infty \rangle = \prod_{i} {c}^{\dagger}_{si} |  0 \rangle$, 
we see that the effect of the operator
$\hat{j}$ is to create kink-antikink pairs of length $d=\pm 1$ at any
of the $m$ possible sites. The matrix elements  with the single pair states
of Eq.(\ref{eq:wavef}) are thus given, in terms
of the eigenfunctions of the semi-infinite chain, 
as
\begin{equation}
  \label{eq:matrixel}
  |\langle \psi_\infty | \hat{j} | \Psi_{K,\lambda}\rangle |^2=
t^2 \frac{L}{s}
\delta_{K,0}  \; | \varphi_{K,\lambda}(1)|^2.
\end{equation}

\subsection{Coulomb potential}

The optical absorption can be divided in two parts
$\sigma=\sigma_{exc}+\sigma_{sc}$, which represent respectively 
the sharp transitions from the ground state to the  
excitonic states and an absorption band due to transitions to the
kink/anti-kink continuum. The corresponding  energies
are obtained from  
Eq.(\ref{eq:bound}) and Eq.(\ref{eq:continuum}) by setting $K=0$.
The excitonic part is an infinite series of delta function peaks 
\begin{equation}
  \label{eq:discr}
  \sigma_{exc} = 
    \sum_{n} 
 \frac{\pi t^2}{s E_{0,n} } {\left| \varphi_{0,n}(1) \right|}^2  
 \delta \left( \omega - E_{0,n} \right) 
\end{equation}
where the spectral weights are given by Eqs. (\ref{wavedisc}) and
(\ref{normdisc}) above, making use of   the property ${}_2F_1=1$ at $d=1$.
Setting $k(\omega)= \arccos \left[ -(\omega - \Sigma_s)/4t \right]$
and converting the sum over states in Eq. (\ref{eq:Kubo}) into an integral,
we obtain the following  analytical expression
for the continuous  absorption band:
\begin{equation}
  \label{eq:optcond}
  \sigma_{sc}(\omega) = 
\left( \frac{\pi V}{4  s^4 \omega} \right) 
\frac{e^{-\displaystyle\frac{V}{2s^3 t}\frac{k(\omega)}{\sin k(\omega)}}}
{1-e^{\displaystyle  -\frac{V}{2s^3 t}\frac{\pi}{\sin k(\omega)}}}
\end{equation}
Note that since we are assuming $t \lesssim V/s^3$, 
the denominator can be set equal to
$1$ up to exponentially small corrections 
(these become important close to the melting of the GWL, 
where anyway the single pair approximation breaks down).

\begin{figure}[htbp]
  \centering
  \resizebox{8cm}{!}{\includegraphics{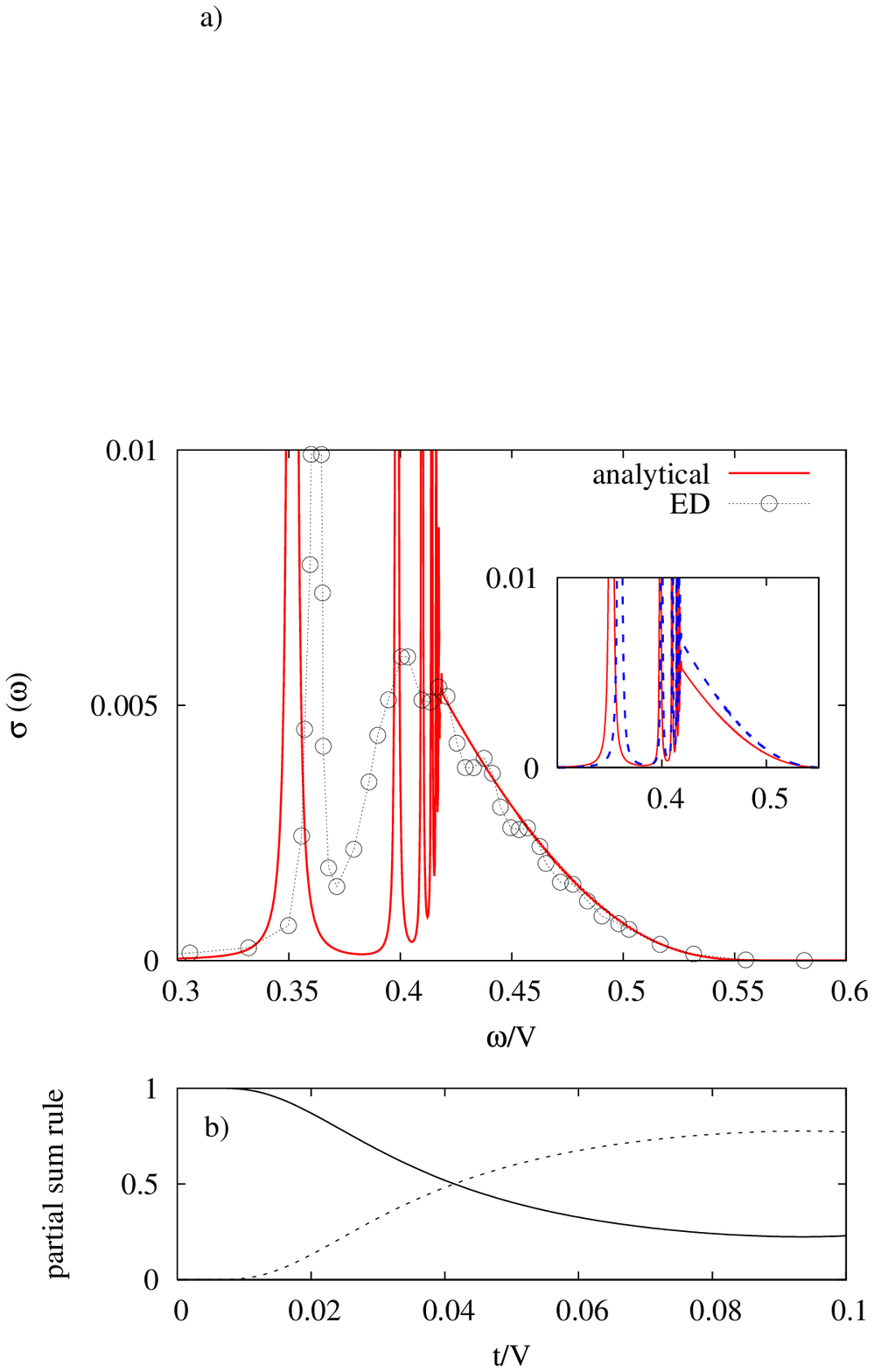}}
  \caption{a) Optical conductivity 
of the generalized Wigner lattice
 in the single pair approximation  
as given by Eqs. (\ref{eq:discr}) and  (\ref{eq:optcond})
 at $n=1/2$
    and $t/V=0.02$. 
    A finite broadening has been
    introduced for graphical purposes. 
The open circles are the  exact
    diagonalization data of Ref. \cite{Mayr}. 
In the inset, the same curve is compared
with the   result obtained with the full potential of
    Eq.(\ref{eq:kk-energy-general}) (dashed lines). \\
b) Fraction of the spectral weight carried respectively 
by the excitons (full
lines) and the kink/anti-kink continuum (dashed line).}
  \label{fig:optcond}
\end{figure}

The optical absorption determined above is  illustrated in Fig. 
\ref{fig:optcond}.a and agrees remarkably well with the exact diagonalization 
results of Ref.\cite{Mayr} The main discrepancy is a shift in the 
position of the first excitonic peak, that 
 can be  entirely ascribed to the the use of the pure
Coulomb potential in Eq. (\ref{eq:pot}). 
A transfer of spectral weight 
from  the excitons to the kink/anti-kink continuum takes place as $t/V$
increases, as shown in Fig. \ref{fig:optcond}.b, where we have plotted
the quantities $\int d\omega \sigma_{exc}(\omega)/\int d\omega
\sigma(\omega)$ and $\int d\omega \sigma_{sc}(\omega)/\int d\omega
\sigma(\omega)$. Note that
formulas (\ref{eq:discr}) and (\ref{eq:optcond}) 
obey the following
property, that can be checked by direct evaluation:
\begin{equation}
  \label{eq:inter}
  \langle \omega \rangle=\int d\omega \sigma(\omega) \omega=\pi t^2/s.
\end{equation}
This relation shows that 
the ``average'' absorption frequency is independent on the
coupling constant and can be used  to  measure directly the value of the 
hopping integral in an optical absorption experiment.

In the above derivation 
 it was assumed that the potential 
follows a pure Coulomb law, Eq.(\ref{eq:pot}), which allowed us to
obtain explicit analytical formulas for the optical absorption. 
However, the discrete Schr\"odinger equation (\ref{eq:disc-Schr})
with the exact potential $\Delta_s (d)$ of Eq.(\ref{eq:kk-energy-general}) 
can be solved to arbitrary accuracy with
small numerical effort. The result is presented in the inset of
Fig. \ref{fig:optcond}.a. We see that the analytical
expression (\ref{eq:optcond})  agrees quite well with the numerical
result. Discrepancies arise 
at very small values of $t/V$, where the excitons become
localized on very short lengthscales, which is where 
the potential $\Delta_s (d)$ sensibly deviates  from the
Coulomb law. 

\medskip

Note that an expression  similar to Eq. (\ref{eq:optcond}) 
was derived by Gallinar\cite{Gallinar93} in the
half-filled case  $n=1$ for electrons with spin, in the limit of a
large on-site repulsion $U$.  
In that case the lowest-lying excitations 
correspond to pairs of doubly occupied/empty sites.\cite{Gallagher}
The energy cost for creating such pairs is set by
$U$, while the long-range tail of the Coulomb potential
acts to bind the opposite charges together.  
\footnote{In the antiferromagnetic phase the problem is formally equivalent to the 
one treated here [set $x=1$ in Eq.(3.1) of Ref.\cite{Gallinar93}]. 
For more general magnetic arrangements, the excitation spectrum must 
be modified to account for the probability of finding neighboring sites 
with opposite spins, leading to a more complex scenario for the optical absorption.}
Similar effects have also been discussed in the context of
conjugated polymers, in which case  the energy scale of an electron-hole
exciton  is set by the Peierls dimerization gap \cite{Abe}
(see also Ref. \cite{Gallagher}).

\subsection{Screened potentials}

The strong asymmetry  of the optical lineshape Eq.  (\ref{eq:optcond}) 
ultimately follows from the long-range nature of the Coulomb potential. 
Indeed, although the energies of the scattering states are given by the
non-interacting expression (\ref{eq:continuum}), 
 their wavefunctions strongly differ from the free plane waves
especially at short distances (the scattering states are required to be
orthogonal to the bound states). 
This is reflected in the optical absorption, which involves 
the wavefunction precisely at $d=1$, where the condition of 
orthogonality with the excitonic states
is most stringent.  To illustrate this issue, we have considered 
the case of a screened potential 
$V/d\to  V \exp\left(- d / \ell \right) /  d$. 
This potential 
is convex and ensures the stability of the  classical 
configuration considered here,\cite{Hubbard78} 
which is
preferable to the case of arbitrarily truncated Coulomb interactions,
for which the ground state is not always univocally determined.

\begin{figure}[htbp]
  \centering
  \resizebox{8cm}{!}{\includegraphics{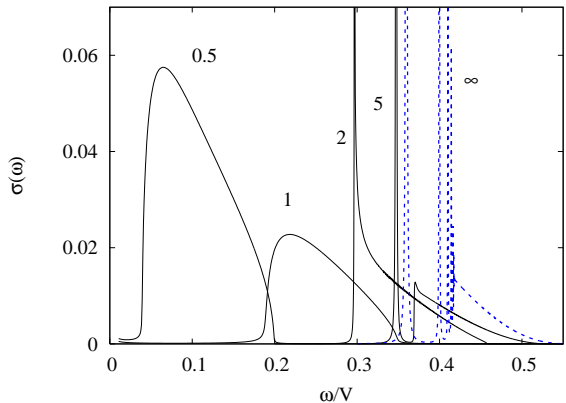}}
  \caption{Optical absorption for a screened potential at $n=1/2$ and $t/V=0.02$.  The labels
    indicate the value of the screening length $\ell$. The dashed line
    corresponds to the unscreened Coulomb potential. }
  \label{fig:screened}
\end{figure}

The results obtained for generic values of the screening length 
 are reported in Fig. \ref{fig:screened} at filling $n=1/2$, for $t/V=0.02$. 
Upon reducing the screening length, the optical absorption shifts to
lower frequencies, the asymmetry of the lineshape becomes less marked  
and the excitonic peaks progressively
disappear (below a critical value, no bound states are possible  and
$\sigma_{exc}=0$).   
In the limit $\ell\to 0$, a truncated nearest-neighbor potential is recovered. 
In that case the effective interaction $\Delta(d)=V e^{-1/\ell}\equiv V_1$ 
is constant for all $d$  
and the scattering states reduce to free-particle states. 
The corresponding current-current correlation function 
has a semi-circular shape and the resulting absorption band is
\begin{equation}
  \label{eq:opt-nn}
  \sigma_{nn}(\omega) = \frac{t}{2\omega} 
\sqrt{1-\left( \frac{\omega- V_1}{4t}\right)^2}. 
\end{equation}

To conclude this section, let us stress that 
the experimental observation of an asymmetric lineshape of
the form shown in Fig.\ref{fig:optcond}.a can be taken as a signature of the
importance of long-range electron-electron interactions in a given
material, and calls for theoretical models that go beyond the intersite
repulsion term usually considered.\cite{Seo}

\section{Spectral function at half filling}

In a photoemission experiment, an electron is added or removed from
the system in a sudden process that is described by the single particle 
spectral function $A(k,\omega) = - \frac{1}{\pi} \mbox{Im} 
G \left( k,\omega \right)$. 
In the spirit of the strong coupling approach presented in the
preceding section,  adding an electron 
to the classical configuration at $n=1/s$ creates $s$ kinks of charge
$1/s$. Analogously, removing an electron creates $s$ anti-kinks of
negative charge $-1/s$. In the special case $s=2$, two defects are
created and the problem becomes equivalent to the one treated
previously, albeit with repulsive interactions 
(the analogy  is no longer true for larger values of $s$, where
 the creation or removal of an electron take the system away from 
the single-pair subspace). 

Classically, since the interaction potential is  repulsive,  
the  state of lowest energy in the system with one particle added (removed)
corresponds to configurations where the two kinks (antikinks) are far
apart. The creation energy of each kink is $\Sigma_2/2=V/4$ 
(cf. Eq. (\ref{eq:pot})).

In the quantum case, we can write the 
1D-Schr\"odinger equation for the two-body
problem, which is formally identical to the one presented in
Sec. II,  except for a change of sign in the interaction potential, 
which results in a symmetrically reflected excitation spectrum: 
the low energy part  now corresponds to the domain-wall continuum,
while the discrete levels are moved to higher energies. 
Such \textit{anti-bound} states  form, as for the excitonic states 
encountered previously, due to the long-range nature of the repulsive 
potential. 

\begin{figure}[htbp]
\resizebox{8cm}{!}{\includegraphics{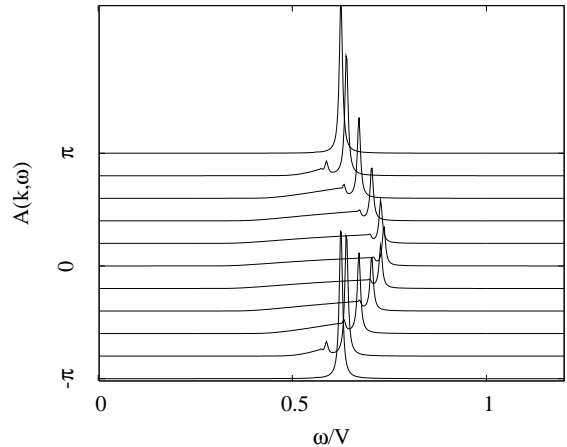}}
  \caption{Spectral function $A(k,\omega)$ for $s=2$ and
    $t/V=0.05$. The different curves correspond to different values of
    $k$, as  indicated on the left axis. }
  \label{fig:Akw}
\end{figure}

The spectral function for $\omega>0$, which describes the inverse
photoemission processes, 
can be calculated  as
\begin{equation}\label{Akw}
  A(q,\omega)=\sum_{K,\lambda} |\langle \Psi_{K,\lambda} | c^+_q
  |\psi_{GS}\rangle|^2 \delta(\omega-E_{K,\lambda})
\end{equation}
where $c^+_q=\frac{1}{\sqrt{L}}\sum_m e^{iqm} c_m^+$ and
$|\Psi_{K,\lambda}\rangle$ are the solutions of the repulsive Schr\"odinger
equation  (\ref{eq:disc-Schr}). 
An analogous formula holds for the spectral function at $\omega<0$,
with $A(q,-\omega)=A(q,\omega)$.
Replacing as usual the ground state with the classical configuration
at $t=0$ enforces the constraint that 
electrons can only be created on the unoccupied sites. 
The resulting expression for the spectral function is
\begin{equation}\label{Akw-result}
  A(q,\omega)=\frac{1}{2}\sum_{\lambda} |\varphi_{q,\lambda}|^2 \delta(\omega-E_{q,\lambda})
\end{equation}
where the sum runs over both the continuous spectrum ($\lambda\equiv
k$) and the discrete anti-bound states ($\lambda\equiv n$), whose
energies are given by  Eqs. (\ref{eq:continuum}) and (\ref{eq:bound}),
that we reproduce here
(in the latter, the quantity $E-\Sigma_s$ changes sign 
due to the repulsive nature of the interactions):
\begin{equation} \label{disp}
  E_{q,n}=\frac{V}{2}+\sqrt{16 t^2 \cos^2(q)+\left(\frac{V}{8(n+1)}\right)^2}
\end{equation}

The results are reported in Fig. \ref{fig:Akw}. The width of the continuous
spectrum is modulated by $q$ and vanishes at $q=\pm\pi$. 
The anti-bound states  are clearly observable as well defined  
quasi-particle peaks in $A(k,\omega)$, that follow the roughly
sinusoidal dispersion relations Eq.(\ref{disp}).

\section{Discussion and Conclusions}

In this work, we have examined the spectral properties of spinless 
electrons  interacting through the long-range Coulomb potential  
on a one-dimensional lattice, which  constitutes a minimal model 
to address the effects of electronic correlations in narrow-band
solids, away from the most studied half-filled case. 
In the strongly interacting regime, the charges form a generalized Wigner
lattice, whose elementary excitations are fractionally charged
domain-walls, themselves interacting through long-range forces. 
Taking advantage of the exact solution of the
Coulomb problem on a one-dimensional chain, 
we have derived an analytical expression for the optical conductivity 
at simple commensurate fillings $n=1/s$, 
as well as for the single particle spectral function in the special case
$n=1/2$. Both are in good agreement 
with the available exact diagonalization
data at $n=1/2$.\cite{Mayr,Daghofer}

The  sharp peaks emerging in the 
optical conductivity, signaling the formation of domain-wall excitons,
the asymmetric lineshape of the absorption continuum, 
as well as the sharp quasi-particle peaks in
the single particle spectral function, 
are all robust features that follow from the
long-range nature of the Coulomb potential, and that are lost in
models with strongly screened or truncated interactions.
These features should be clearly observable in experiments, allowing to address
the relevance of long-range interactions in the charge-ordered insulating 
phases of narrow-band one-dimensional systems.  

As an example, the mid-infrared 
optical absorption band recently measured\cite{Yamamoto} in the organic
quarter-filled ($n=1/2$) 
salt (DI-DCNQI)$_2$Ag --- a prototypical system
exhibiting Wigner crystal ordering\cite{Itou} --- resembles very closely the
theoretical spectra obtained from Eq.  (\ref{eq:discr}) and (\ref{eq:optcond}),
and a satisfactory fit can be obtained with physically sound 
parameters ($V=1.1eV$ and $t=80meV$).
Absorption shapes similar to the ones presented in this work have also
been reported  in other classes of organic compounds, such as
DBTSF-TCNQF$_4$, TTF-TCNQ\cite{Jacobsen} 
and Cs$_2$(TCNQ)$_3$,\cite{Cummings} 
corresponding respectively to $n=1$, $n\simeq 1/2$ and $n=1/3$, whose
detailed analysis is postponed to future work.

\section*{Acknowledgments}
For one of us, G.R., this work was funded by the Swiss-Italian
foundation  ``Angelo Della Riccia''. S.F. gratefully acknowledges enlightening
discussions with D. Baeriswyl.


\begin{thebibliography}{99} 
 
\bibitem{Voit} J.\ Voit, Rep.\ Prog.\ Phys.\ {\bf 57}, 977 (1994).

\bibitem{Rice} M.\ J. Rice, A.\ R.\ Bishop, J.\ A.\ Krumhans, S.\ E.\ Trullinger, Phys.\ Rev.\ Lett.\ {\bf 36}, 432 (1976).

\bibitem{Laughlin} R.\ B.\ Laughlin, Rev.\ Mod.\ Phys.\ {\bf 71}, 863 (1999).

\bibitem{Hubbard78} J.\ Hubbard, Phys.\ Rev.\ B {\bf 17}, 494 (1978).



\bibitem{Itou} K. Hiraki and K. Kanoda
Phys. Rev. Lett. 80, 4737-4740 (1998); 
T.\ Itou, K.\ Kanoda, K.\ Murata,  T.\ Matsumoto, K.\ Hiraki, T.\ Takahashi, Phys.\ Rev.\ Lett.\ {\bf 93} 216408 (2004). 

\bibitem{Abbamonte} P.\ Abbamonte et al., Nature {\bf 431}, 1081
  (2004); A. Rusydi et al., Phys. Rev. Lett {\bf 97}, 016403 (2006)

\bibitem{Horsch} P.\ Horsch, M.\ Sofin, M.\ Mayr, and M.\ Jansen, Phys.\ Rev.\ Lett.\ {\bf 94} 076403 (2005).

\bibitem{Valenzuela} B.\ Valenzuela, S.\ Fratini, and D.\ Baeriswyl, Phys.\ Rev.\ B {\bf 68} 045112 (2003).

\bibitem{SynthMet} S.\ Fratini, B.\ Valenzuela and D.\ Baeriswyl, Synt. Met. {\bf 141}, 193 (2004).

\bibitem{Mayr}     M.\ Mayr, P.\ Horsch, Phys.\ Rev.\ B {\bf 73} 195103 (2006).

\bibitem{Gallinar84} J.-P.\ Gallinar, Phys.\ Lett.\ {\bf 103}A, 72 (1984).

\bibitem{Kvitsinski} A.\ A.\ Kvitsinski, J.\ Phys.\ A: Math.\ Gen.\ {\bf 25}, 65 (1992).


\bibitem{Daghofer}  M.\ Daghofer, P.\ Horsch, cond-mat/0608062 (2006).  

\bibitem{Slavin} V. Slavin, Phys. Stat. Sol. {\bf 242} 2033 (2005)

\bibitem{Barford} W.\ Barford, Phys.\ Rev.\ B {\bf 65}, 205118 (2002).

\bibitem{Gallinar93} J.-P.\ Gallinar, Phys. Rev. B {\bf 48}, 5013 (1993).

\bibitem{Gallagher} F.\ B.\ Gallagher, S.\ Mazumdar, Phys.\ Rev.\ B {\bf 56}, 15025 (1997).


\bibitem{Abe} S.\ Abe, J.\ Phys.\ Soc.\ Japan {\bf 58}, 62 (1989).

\bibitem{Seo} H. Seo, J. Merino, H. Yoshioka, M. Ogata, 
J. Phys. Soc. Jpn. {\bf 75}, 051009 (2006)

\bibitem{Yamamoto} K. Yamamoto, T. Yamamoto, K. Yakushi, C. Pecile,
  M. Meneghetti, Phys. Rev. {\bf B 71} 045118 (2005) 

\bibitem{Jacobsen} C.\ S.\ Jacobsen, I.\ Johanssen, K.\ Bechgaard, Phys.\ Rev.\ Lett.\  {\bf 53}, 194 (1984).

\bibitem{Cummings} K.\ D.\ Cummings, D.\ B.\ Tanner, J.\ S.\ Miller, Phys.\ Rev.\ B {\bf 24}, 4142 (1981).

\end{thebibliography}
\end{document}